\documentclass[graphicx]{JHEP}
\usepackage{epsfig,amssymb}
\usepackage{ulem}

\newcommand{\bmat}{\left(\begin{array}}
\newcommand{\emat}{\end{array}\right)}
\newcommand{\beq}{\begin{equation}}
\newcommand{\eeq}{\end{equation}}
\newcommand{\wt}{\widetilde}

\def\ra{\rightarrow}

\def\f{\frac}

\def\U{$U(1)_X$\,}
\def\L{\left(}

\def\bwt{\begin{widetext}}
\def\ewt{\end{widetext}}
\def\be{\begin{equation}}
\def\ee{\end{equation}}
\def\bea{\begin{eqnarray}}
\def\eea{\end{eqnarray}}
\def\bean{\begin{eqnarray*}}
\def\eean{\end{eqnarray*}}
\def\bary{\begin{array}}
\def\eary{\end{array}}
\def\bit{\begin{itemize}}
\def\eit{\end{itemize}}

\def\lan{\langle}
\def\ran{\rangle}

\def\ra{\rightarrow}

\def\ld{\lambda}

\def\su5u1{SU(5) \times U(1)}
\def\fsu5u1{SU(5) \times U(1)'}
\def\so10{SO(10)}
\def\sq20{SO(10) \times SO(10)}
\def\R{\right)}
\def\rw{\rightarrow}

\def\Ld{\Lambda}
\def\ld{\lambda}

\def\f{\frac}

\def\bwt{\begin{widetext}}
\def\ewt{\end{widetext}}
\def\be{\begin{equation}}
\def\ee{\end{equation}}
\def\bea{\begin{eqnarray}}
\def\eea{\end{eqnarray}}
\def\bean{\begin{eqnarray*}}
\def\eean{\end{eqnarray*}}
\def\bary{\begin{array}}
\def\eary{\end{array}}
\def\bit{\begin{itemize}}
\def\eit{\end{itemize}}

\usepackage[centertags]{amsmath}
\usepackage{amssymb}

\title{Light Dark Matter from the $U(1)_X$ Sector in the NMSSM with Gauge Mediation}

\author{Zhaofeng Kang$^1$, Tianjun Li$^{1,2}$, Tao Liu$^1$, Chunli Tong$^1$, Jin Min Yang$^1$ \\
$^1$ Key Laboratory of Frontiers in Theoretical Physics,
         Institute of Theoretical Physics, Academia Sinica,
         Beijing 100190, P. R. China\\
$^2$ George P. and Cynthia W. Mitchell Institute for Fundamental Physics,
     Texas A$\&$M University, College Station, TX 77843, USA }

\abstract{Cosmic ray anomalies observed by PAMELA and Fermi-LAT
experiments may be interpreted by heavy (TeV-scale) dark matter
annihilation enhanced by Sommerfeld effects mediated by a very
light (sub-GeV) $U(1)_X$ gauge boson, while the recent direct
searches from CoGeNT and DAMA/LIBRA experiments may indicate a
rather light ($\sim 7$ GeV) dark matter with weak interaction.
Motivated by these apparently different scales, we consider a
gauge mediated next-to-the minimal supersymmetric standard model
(NMSSM) entended with a light \U sector plus a heavy sector $(\bar
H_h,H_h)$, which can provide both a light ($\sim 7$ GeV) and a
heavy (TeV-scale) dark matter without introducing any ad hoc new
scale. Through the Yukawa coupling between $H_h$ and the messager
fields, the \U gauge symmetry is broken around the GeV scale
radiatively and a large negative $m_S^2$ is generated for the
NMSSM singlet $S$. Furthermore, the small kinetic mixing parameter
between \U and $U(1)_Y$ is predicted to be $\theta\sim
10^{-5}-10^{-6}$ after integrating out the messengers. Such a
light dark matter, which can have a normal relic density from the
late decay of the right-handed sneutrino (assumed to be the
ordinary next-to-the lightest supersymmetric particle and
thermally produced in the early Universe), can serve a good
candidate to explain the recent CoGeNT and DAMA/LIBRA results.}

\begin{document}
\maketitle  \indent

\newpage
\section{Introduction}
The recent indirect dark matter detection experiments like PAMELA \cite{pamela}
and Fermi-LAT \cite{Abdo:2009zk} found  cosmic ray anomalies,
which can be interpreted by dark matter annihilation or decay
(although some astrophysical explanations like pulsars are also possible).
This inspires the construction of a class of models with a light dark
\U sector \cite{ArkaniHamed:2008qn}, which gives a sub-GeV dark gauge boson.
Such a sub-GeV gauge boson plays a key role in the dark matter explanation
of the cosmic ray anomalies: for the annihilating dark matter it can induce
large Sommerfeld enhancement and kinetically forbid the hadronic products from
the annihilation, while for the decaying dark matter \cite{Arvanitaki:2008hq}
it can suppress the hadronic activity \cite{Ruderman:2009tj}.
At the same time, some dark matter direct detection experiments
such as DAMA/LIBRA \cite{Bernabei:2010mq}, CDMS II \cite{Ahmed:2009zw}
and CoGeNT \cite{Aalseth:2010vx} also reported some plausible evidence
of dark matter, which, including the null result from XENON10 \cite{Angle:2007uj},
may be accommodated by a quite light dark matter at GeV scale ($\sim 7$ GeV)
with a dark matter nucleon scattering cross section $\sigma_p\sim 10^{-40}$
cm$^{2}$ \cite{Hooper:2010uy}. This has inspired some recent studies on the light dark
matter \cite{Fitzpatrick:2010em,Essig:2010ye}.

With Sommerfeld enhancement, it seems to us that
 the dark matter explanation for all these experiments must
involve three very different scales: the TeV-scale heavy dark matter
(HDM), the GeV-scale light dark matter (LDM), and the sub-GeV $U_X(1)$ dark sector.
It is then quite challenging to embrace all these aspects in one framework.
Firstly, it is not a trivial problem to accommodate such a light \U gauge boson
at low energy without introducing a new scale by hand.
As is well known, supersymmetry (SUSY) helps to stabilize a scale and, moreover,
its breaking usually generates a new scale which is encoded in the soft
SUSY breaking terms. Thus the crucial task is to obtain a proper
GeV-scale soft Lagrangian for the Higgs fields in the \U sector.
As proposed in \cite{ArkaniHamed:2008qn} and then followed in
\cite{Chun:2008by,Baumgart:2009tn,Cheung:2009qd,Morrissey:2009ur},
SUSY breaking (maybe exhibited as soft terms) in some hidden sector may be
gauge mediated to the \U sector to generate the GeV-scale.
Secondly, although it is not difficult to construct a GeV-scale \U sector
while allows for a sub-GeV gauge boson through introducing a very weakly
charged Higgs field (say $Q_Hg_X\sim 0.03$), the \U sector with such a light
gauge boson will usually also predict some other Higgs bosons as light as
the gauge boson and the LDM annihilates to these bosons very effectively,
leading to a very small relic density after freezing out (say
$\Omega_{LDM}h^2\sim 10^{-4}$). Some studies \cite{Essig:2010ye} showed that
even with such a small relic density the LDM may still generate scattering
signals at the dark matter detectors if the LDM-quark coupling strength
is enhanced enough. Nevertheless, it would be more natural if the LDM density
is at a normal level ($\sim 0.1$). Thus, the LDM may be understood to
be mainly produced from the late decay of the ordinary next-to-the
lightest sparticle
(ONLSP) in the visible sector. This may be a reasonable conjecture since in the
presence of some new light $R-$odd state in the \U sector, the ONLSP may
decay to this sector with proper time scale.

In this work we try to extend the gauge mediated
next-to-the minimal supersymmetric standard model (NMSSM) with
a light \U sector plus a heavy sector $(\bar H_h,H_h)$,
which can provide both a light ($\sim 7$ GeV) and a heavy (TeV-scale)
dark matter without introducing any ad hoc new scale.
In our framework the crucial dynamics is that the HDM couples directly
through Yukawa couplings with the messenger fields which carry
the \U charge. The \U gauge symmetry can be broken around the
GeV scale radiatively, and a large negative $m_S^2$
is generated for the NMSSM singlet $S$. Interestingly, the
small kinetic mixing parameter between \U and $U(1)_Y$
is predicted to be $\theta\sim 10^{-5}-10^{-6}$ after integrating out
the messenger fields. Such a light dark matter, which can have a normal relic
density from the late decay of the right-handed
sneutrino, can be a good
candidate to explain the recent CoGeNT and DAMA/LIBRA data.

This work is organized as follows. In Section II we present the
model. In Section III we discuss its concrete realization.
Finally,  discussions and  conclusion are given in Section IV.
In Appendix A, we explain the kinetic mixing and dark-visible interaction.
In Appendix B, we present the soft terms from HDM-messenger direct
couplings. And in Appendix C, we give the one-loop renormalization
group equations (RGEs) of some soft terms.

\section{Model Building}
Our model  is based on the NMSSM with gauge mediated SUSY breaking
(GMSB). And it has two features: (1) The NMSSM singlet $S$
naturally provides a TeV scale to explain the origin of the HDM
mass scale; (2) Through radiative correction with $1/16\pi^2$
suppression, the GMSB provides a simple way to generate the
GeV-scale for the \U dark sector. Some previous studies on this
line have been carried out
\cite{ArkaniHamed:2008qn,Chun:2008by,Baumgart:2009tn,Cheung:2009qd,Morrissey:2009ur}.
In our study we will intensively examine the dark matter
phenomenology in the NMSSM extended with the $U(1)_X$ sector and
the extra TeV-scale degree of freedoms, paying special attention
to the mechanism of the $U(1)_X$ breaking at GeV-scale. We find
that if the conventional hidden sector messengers are slightly
charged under \U, then the soft terms in the \U dark sector can be
at a proper scale. Our work will address the following problems in
a coherent framework:

\subsection{Generating a Large Negative Soft Mass-Square for $S$}
As a simple extension of the MSSM, the NMSSM \cite{NMSSM} can
solve the $\mu$ problem and the little hierarchy problem
\cite{Dermisek:2005ar}, and thus has recently attracted much
attention \cite{NMSSM-pheno}. However, in the mechanism of the
GMSB it is difficult to construct a phenomenologically acceptable
NMSSM \cite{Dine:1993yw,deGouvea:1997cx}. The key difficulty is
that the singlet $S$ couples only to the Higgs doublets and thus
the soft term $m_S^2$  can not be generated negative enough at the
weak scale through RGE. To solve this problem, some efforts have
been made, e.g., coupling $S$ to extra $SU(3)_C$-charged particles
\cite{Dine:1993yw} or directly to messengers
\cite{Giudice:1997ni,Chacko:2001km,Delgado:2007rz}. In our
framework, since we have extra states ($\bar H_h,H_h$) which
couple to $S$, we can obtain large enough $m_S^2$ by only coupling
$H_h$ or $\bar H_h$ directly to messengers ($S$ does not couple to
messengers). In fact, this is a natural choice since this coupling
leads to a large ($\sim $TeV$^2$) splitting between the soft
mass-square $m_{\bar H_h}^2$ and $m_{H_h}^2$ at the messenger
boundary. That significantly impacts on the evolution of the soft
mass-square of the dark Higgs field, leading to a negative
mass-square and breaking the \U in the dark sector. The dynamics
of this part is described by the superpotential
\begin{align}
W_{1}= &\L\lambda SH_uH_d+{\kappa \over 3}S^3\right)+\lambda_h S\bar
H_hH_h\cr & +\bar H_h\L \ld_T T_1\bar T_2+\ld_DD_1\bar D_2\R+X\L
\xi_{1,T}\bar T_1T_1+\xi_{1,D}\bar D_1D_1 +(1\rw 2)\R,
\end{align}
where $X$ is the spurion Goldstino field parameterized as
$X=M+\theta^2 F$, and $(T_i,D_i)=f_i$ and $(\bar T_i,\bar D_i)=\bar
f_i$ form $(5,\bar 5)$ representation of $SU(5)-$GUT group.  The
matter and messenger fields have the assignments under the $Z_3-$symmetry
of the NMSSM and the \U :
\begin{align}
&S\rw e^{i\pi/3}S,\quad H_h\rw e^{i2\pi/3}H_h,\quad \bar H_h\rw \bar
H_h, \cr &[f_1]=-[\bar f_1]=Q_{f_1},\quad [\bar
H_h]=-[H_h]=-Q_{H_h},
\end{align}
while all other fields are neutral under the above symmetries, thus
$Q_{f_1}=Q_{H_h}$.

Let us comments on the superpotential:
\begin{itemize}
\item[(1)] The superpotential has a $Z_2^h$ symmetry to keep the HDM
stable (the messengers $(\bar f_1,f_1)$ are $Z_2^h-$odd).
According to a recent study~\cite{Feng:2010zp}, the explanation of PAMELA
through such HDM annihilation with Sommerfeld enhancement is difficult.
In particular, the maximal Sommerfeld enhanced factor is about 100 for a TeV
scale heavy dark matter. To explain the PAMELA and Fermi-LAT experiments,
for simplicity, we assume that the dark matter
 density in the sub-halo is about three or four times larger than
the usual value. By the way,
 to explain PAMELA, we had better resort to decaying HDM.
To let our HDM to decay to dark gauge bosons, we need to break the
$Z_2^h$ symmetry by introducing some new mechanism~\cite{Ruderman:2009tj}.
We will not further discuss the HDM phenomenology in this work.
Instead, we will focus on the LDM phenomenology.

\item[(2)] It is important to arrange \U charge to forbid coupling like
$\lambda_f\bar H_h \bar f_i f_i$, which leads to a one-loop  tadpole
for $S$ in the superpotential after integrating out the messengers:
$\int d^2\theta \xi S $ with $\xi \sim\lambda_f^2 F/16\pi^2$, which
tends to destabilize the weak scale. But at the messenger boundary,
a large negative mass-square for $S$ is generated at two loop as
\begin{align}\label{ms2}
m_S^2=&-\f{1}{(16\pi^2)^2}\L3\ld_T^2+2\ld_D^2\R\ld_h^2\f{F^2}{M^2},
\end{align}
which can be as large as several-hundred GeV, depending on the couplings.
For example, for $M\simeq10^8$ GeV at the messenger boundary
and taking Yukawa couplings as $\ld_h\sim 1$,
$\ld_T\simeq\ld_D\sim0.2$,  we have $m_S^2\sim -(280$ GeV)$^2$.
In this way, it is possible to make the NMSSM in the GMSB to have
successful electroweak symmetry breaking.
\item[(3)] The Yukawa coupling $\ld_h$ plays an important role.
In additional to generate a large $m_S^2$, a large $\ld_h$ is also
required for having a HDM. From Eq.~(\ref{ldh}), the one-loop
evolution of $\ld_h$ below the messenger scale is approximated as
(drop the small contribution from  $\ld,\kappa$ and $Q_hg_X$)
\begin{align}
\ld_h(M_{susy})\approx \L\f{1}{
\ld_h^3(M)}-\f{18}{16\pi^2}\log\f{M_{susy}}{M}\R^{-1/3}.
\end{align}
We need $\ld_h(M_{susy})\sim 0.5-1$ (depending on the value of $v_s$)
to have a heavy HDM. Besides, it makes the HDM to annihilate to some states
in the NMSSM effectively so that to have small relic density.
In this way the HDM can avoid direct detection and explain the cosmic ray anomaly
by a proper shorter lifetime than the decaying HDM with the assumption
$\Omega_{HDM}h^2\simeq 0.12$.
\end{itemize}

\subsection{Generating  A Small Negative Soft Mass-Square for the Dark Higgs $H$}
In our model we assume that the dark sector respects a global
$SU(N)$ flavor symmetry (it can be gauged to form a non-Abelian dark
sector \cite{Chen:2009ab,Baumgart:2009tn}, but in this work we do
not discuss this case). This symmetry is useful because it can
protect the light dark matter candidate to be stable and allow to
construct a simple dark sector without anomaly if we require the
dark sector has no \U singlet (we will explain why we do not prefer
a singlet later). The minimal field content includes: $(\bar
H_l,H_l)$ carrying \U charge $(Q_{\bar H_l},Q_{H_l})$ and forming
the $(\bar N,N)$ representation of $SU(N)$; the dark Higgs $H$
carrying \U charge $Q_{H_h}$ and being a flavor singlet. Then, under
the symmetry \U$\times SU(N)$, the most general superpotential takes
a very simple form:
\begin{align}\label{dark}
W_{dark}=\lambda_l H\bar H_l H_l.
\end{align}
The $U(1)_X^3$ anomaly cancellation and the \U neutral condition
lead to two equations:
\begin{align}\label{}
Q_{H_h}+ N (Q_{H_l}+Q_{\bar H_l})=0,\quad
Q_{H_h}^3+N(Q_{H_l}^3+Q_{\bar H_l}^3)=0.
\end{align}
Note that other \U-charged states in our model are vector-like, and thus
do not contribute to anomaly. Especially, it has a
nontrivial solution
\begin{align}\label{charge}
Q_{H_l}=-{Q_{H_h}\over {2N}}\L1-{\rm
sign}(Q_{H_h})\sqrt{\frac{4N^2-1}{3}} \R,
\end{align}
\begin{align}
Q_{\bar H_l}=-{Q_{H_h}\over {2N}}\L1+{\rm
sign}(Q_{H_h})\sqrt{\frac{4N^2-1}{3}} \R.
\end{align}
Another solution is trivial, obtained by exchanging the role of
$\bar H_l$ and $H_l$. For any allowed $N$, $Q_{H_l}$ and $Q_{\bar
H_l}$ take opposite sign with $Q_{H_h}$. This is a required property
to assure that only $H$ gets negative soft mass-square.

>From the requirement of a negative $m_{H}^2$ at the dark scale
$\mu_d$, $Q_{H_h}$ is determined to take the same sign with
$Q_{H_h}$. For pure GMSB, at the messenger boundary, due to
anomaly cancellation, there is a sum rule for the soft terms:
${\cal S}_X\equiv{\rm Tr}(Q_i m_i^2)=0$, with the trace running
over all \U-charged fields. But the \U-charged HDM directly
couples to the messengers in the hidden sector and acquires a
large boundary value through Yukawa mediation (see Eqs.~(\ref{Hh})
and (\ref{bHh})), which violates this sum rule. Consequently, the
non-vanishing ${\cal S}_X$ drastically changes the renormalization
of the dark Higgs soft mass-squares, driving some of them negative
at $\mu_d$. The trace term is then given by
\begin{align}
{\cal S}_X&={\rm Tr}(Q_i m_i^2)=Q_{H_h}(m_{H_h}^2-m_{\bar H_h}^2)\cr
 &\sim Q_{H_h}  {\ld_T^2\over
(16\pi^2)^2}\left[ 16g_3^2(M)-5\ld_h^2(M)\right] {F^2\over M^2}.
\end{align}
The above estimation is based on the requirement that at the scale
$M$, $\ld_{T,D}\ll g_3,\ld_h$, where $g_3(M)\simeq0.9$. There is a
substantial cancellation between the terms in the bracket and thus
in the following estimation we set $16g_3^2(M)-5\ld_h^2\equiv C_T
g_3^2(M)$ with $C_T\sim 1$. Consequently, it generates the low
energy value of $m_{H}^2$ (see Eq.~(\ref{mH2})):
\begin{align}\label{mhl}
m_{H}^2(\mu_d)&\simeq m_{gauge}^2(M)+{2 Q_{H_h} g_X^2\over
16\pi^2}{\cal S}_X\times \log { \mu_d\over M},
\end{align}
where the first term is  the  small soft term contributed by pure
\U gauge  mediation. Then, we can parameterize the low energy
dark Higgs parameter as
\begin{eqnarray}\label{mhl}
m_{H}^2(\mu_d)&\approx& \left[0.16\L {Q_{f_1} g_X\over 0.01}\R^2 \L
{Q_Hg_X\over 0.02}\R^2 \right. \nonumber \\
&& \left. -0.2\L{Q_Hg_X\over 0.02}\R \L{Q_{H_h}g_X\over
0.01}\R \L{\lambda_T\over 0.2}\R^2 \right]{\,\rm GeV^2}.
\end{eqnarray}
Here we set $F/M=10^5$ GeV, $C_T=0.5$ and a typical  dark scale
$\mu_d\sim 10$ GeV. With a moderate arrangement for $Q_{f_1}\,g_X$
and $\ld_T$, we readily get $0>m_{H}^2(\mu_d)\gtrsim -1$ GeV$^2$.
Note that we do not need cancellation between the two
contributions. In practice, we only require that the second term
dominates over the first term and takes small value.  The soft
mass-square for $(\bar H_l,H_l)$ can be obtained similarly, which
is enhanced by the trace term because their \U charge is opposite
to $H$.

Let us comment on the above charge assignments. First,
$Q_{f_1}g_X\sim 0.01$ not only determines the soft mass scale from
\U mediation, but also directly relates with the value of $\theta$
discussed later. Next, a small value $Q_{H_h} g_X\sim 0.01$ ont only
helps to make the HDM to avoid direct detection, but also avoid the
unnecessary enhancement by dark gauge boson, which will subject to
the gamma ray constraint \footnote{In principle, we can increase
$Q_hg_X\sim 0.5$ and meanwhile set a smaller value $\ld_T\sim 0.03$.
It makes the HDM still be a candidate for annihilating dark matter
with Sommerfeld enhancement.}. As for the small $Q_Hg_X\sim 0.02$,
controlling the quartic term from D-term, is necessary  to generate
a larger VEV of the dark Higgs, providing the several-GeVs scale for
the  light dark matter.

\subsection{Predicting a Small Kinetic Mixing Parameter $\theta$}
Since the messengers also carry \U charges, our framework naturally predicts
a value for the kinetic mixing parameter $\theta$ between \U and $U_Y(1)$:
\begin{align}
\theta \sim\sum_I n_I {g_1g_XQ_Y^IQ_X^I\over
16\pi^2}\log{M_{GUT}\over M},
\end{align}
where $n_I$ is the number of fields $I$ that carry hypercharge
$Q_Y^I$ and \U charge $Q_X^I$. At first,  the contribution from a
complete representation of $SU(5)$ cancels exactly due to the traceless
generators of $SU(5)$. For example, for $(\bar f_1,f_1)$ we have
\begin{align}
2\times \L3\times {1\over 3} \times Q_{f_1}+2\times\L-{1\over
2}\R\times Q_{f_1}\R=2Q_{f_1}{\rm Tr}(T_{24})=0,
\end{align}
with $T_{24}$ a generator of $SU(5)$ that defines hypercharge. In
general, the small doublet-triplet splittings between messenger
fields can be obtained after the $SU(5)$ gauge symmetry breaking via
the dimension-5 operators $X{\bar f}_i \Phi f_i/{\Ld} $ (with
operator coefficient set to be 1) where $\Phi$ is the ${\mathbf 24}$
representation Higgs field and $\Ld$ is the fundamental scale of
the UV theory such
as string scale $M_{string}\simeq 3.0\times 10^{17}$ GeV, or reduced
Planck scale $M_{\rm Pl}\simeq 2.4\times10^{18}$
GeV~\cite{Li:2010xr}. Note that the $SU(5)$ unification scale is
about $2.4\times 10^{16}$ GeV, thus if we take $\Ld=M_{string}$, we
obtain $|\xi_{1, D}-\xi_{1,T}|/\xi_{1,D}\sim 0.1$.
 In addition, the $\theta$ parameter is given by
\begin{align}\label{theta}
\theta \simeq 2Q_{f_1} {g_1g_X\over 16\pi^2}\log{\xi_{1,T}\over
\xi_{1,D}}\sim {\cal O}(10^{-5})~.~
\end{align}
Thus, we require  $Q_{f_1}g_X\sim 0.05$, which is consistent with the
previous parameterization. By the way, the RGE effects may also
induce the doublet-triplet splitting (see Eqs.~(\ref{xiT}) and
(\ref{xiD})), which could be very small if the corresponding Yukawa
couplings are small. On the other hand, for large Yukawa couplings, proper
splitting is induced even without turning back to the high dimension
operators.



\subsection{Light Dark Matter Candidate}
Recently, dark matter direct detection experiments showed some hints
on light dark matter $\sim 7$ GeV. It is natural to relate it with
the light \U dark sector \cite{Essig:2010ye,Feldman:2010wy}.
Although there is a small gap between the LDM and dark gauge boson
mass scale, it can be explained by a small gauge coupling of the
dark Higgs ($Q_{H} g_X\sim 0.02$), provided that the Yukawa coupling
$\lambda_l$ is about $0.5$. We will elaborate this problem in the
next Section.

\begin{figure}[htb]
\begin{center}
\includegraphics[width=3.6in]{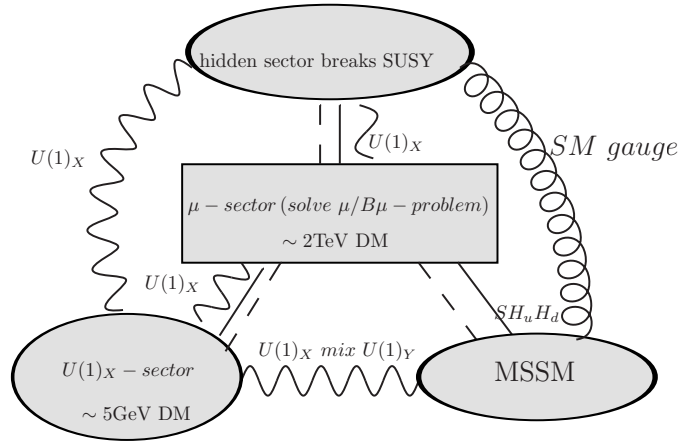}
\end{center}
\vspace*{-1cm} \caption{The schematic diagram showing our dynamics.
Solid and dashed lines denote other possible interactions beyond
gauge interactions.}\label{scheme}
\end{figure}

In summary, we depict our dynamics structure in Fig.~\ref{scheme}.
The hidden sector plays a crucial role in our
framework: it not only generates all the necessary low energy mass
scales, but also explains a small $\theta$ in the dark matter
phenomenology.

\section{Light Dark Matter Phenomenology}

\subsection{Vacuum and Spectra of the Theory}
First we check the vacua of the visible sector and the dark
sector. For the former, the desired electroweak vaccum takes a
form $v_s=\lan S\ran$, $v_u=\lan H_u^0\ran$ and $v_d=\lan
H_{d}^0\ran$. In order to ensure the stability of the HDM and
avoid the breaking of \U at TeV scale, we must have $\lan\bar
H_h\ran=\lan H_h\ran=0$. We need to be cautious about this because
$m_{H_h}^2(M)$ and $m_{\bar H_h}^2(M)$ are negative and take
roughly a value as $m_S^2(M)$ (see Eqs.~(\ref{Hh}) and
(\ref{bHh})). But a negative mass-square does not always mean a
non-zero VEV in the multi-Higgs system. We can prove it by
assuming a vacuum with vanishing $\lan H_h\ran$ and $\lan \bar
H_h\ran$, and then check whether such a vacuum leads to a
tachyonic direction. In practice, in the complex scalar mass
system of $(\bar H_h,H_h^*)$, the mass matrix reads
\begin{align}
M_{H_h}^2=\L\begin{array}{cc}
  (\ld_hv_s)^2+m_{\bar H_h}^2 & (\ld_h v_s)(\kappa\, v_s)+(\ld_h v_s)A_{\ld_h} \\
  (\ld_h v_s)(\kappa\, v_s)+\ld_h v_sA_{\ld_h}&  (\ld_hv_s)^2+m_{ H_h}^2
\end{array}\R.
\end{align}
Obviously, it  is definitively positive provided that
$\ld_hv_s\sim2$ TeV is much larger than other scales in the matrix.
This condition can be satisfied for the following reasons.
First, from the previous parameter estimation, all the soft masses
typically lie much below TeV. Furthermore, to generate a large $v_s$
we require $\kappa\sim 0.1$ ($\ll\ld_h$). Concretely, the lightest
boson has a mass-square approximated by
\begin{align}
M_-^2\approx  (\ld_hv_s)^2+m_{ \bar H_h}^2
-\f{(M_{H_h}^2)_{12}^2}{m_{ H_h}^2-m_{\bar H_h}^2}.
\end{align}
where $(M_{H_h}^2)_{12}$ is the 12-element of $M_{H_h}^2$. This
approximation is valid when $m_{ H_h}^2-m_{\bar
H_h}^2>|(M_{H_h}^2)_{12}|$.  So, in general this is a stable dark
matter in the TeV region required by PAMELA and Fermi-LAT.

We further briefly comment on the pattern of the parameter space and the
symmetry breaking in the NMSSM in our scenario. First, $A_\ld$
and $A_\kappa$ are induced by RGEs, suppressed by loop factor.
Note that a new contribution $A_{\ld_h}\sim -100$ GeV (see
Eq.~(\ref{Aldh})) affects the running of $A_\ld$ and $A_\kappa$
significantly:
\begin{eqnarray}
16\pi^2{dA_\ld\over dt}&\approx&\left(2\ld_h^2A_{\ld_h}+6h_t^2A_t+6g_2^2M_2+...\R,\\
16\pi^2 {dA_\kappa\over dt}&\approx&
\left(2\ld_h^2A_{\ld_h}+12\lambda^2 A_\lambda+...\R.
\end{eqnarray}
Since $\ld_h$ is large in our framework, this new contribution is quite sizable.
Especially, $A_\kappa$ will get a new contribution at order $\sim
-2\f{\L3\ld_T^2+2\ld_D^2\R\ld_h^2}{(16\pi^2)^2}\f{F}{M}\log\f{\mu_d}{M}\sim
{\cal O}(10)$ GeV. As a result, in general one can not expect
a very light $R-$axion (CP-odd) $a$ in the spectrum. But in
case of small $\kappa$ and $\ld$, and $v_s\gg v$,
some parameter space still allows for $m_a< 2m_b$ and consequently
the $R-$axion solution to the fine-tuning problem may be accommodated
\cite{Dermisek:2005ar}. Anyway, these trilinear
terms are small compared with $m_S^2$, and thus the electroweak and
$Z_3$ breaking in the NMSSM is dominantly driven by the negative $m_S^2$.
Approximately, we get a $v_s\gg v$ limit through
\begin{align}
v_s\simeq {m_S\over \kappa}\sim {\cal O}(5){\,\rm TeV},
\end{align}
which is readily achieved by  a small  $\kappa\sim 0.1$.
Incidentally, a small $\kappa$ is a safe choice to stabilize the
HDM mass scale and moreover is favored by keeping the theory
perturbative up to the GUT scale. In conclusion, with the effects
of the Yukawa coupling between the HDM and the messengers, our
scenario is capable of providing a proper solution to the NMSSM.

The \U symmetry breaking and the spectrum in the dark sector
can be analytically studied. The total scalar potential is
$V=V_D+V_F+V_{soft}$, with each term given by \footnote{The study in
\cite{Cheung:2009qd} observes that the effective FI-term
$\xi_I\propto \theta$ generated by  the mixing $D-$term between
$U(1)_Y$ and $U(1)_X$ is able to generate the proper \U breaking,
which is ignored in our study because $\theta$ is small.}
\begin{align}\label{potential}
V_D=&{1\over 2}g_X^2\left[  \left(Q_{H_l} |{  H}_l|^2+Q_{\bar H_l}
|\bar
{  H}_l|^2+Q_{H_h}|H|^2\right)+\xi_X \right]^2,\\
V_F=&
 \left|\lambda_l{  H}_l\bar {
H}_l \right|^2
 + |\lambda_l|^2|H|^2\left(| \bar H_l|^2+| {
H}_l|^2\right),\\
V_{soft}=& m_{H_l}^2|{H_l}|^2+ m_{\bar H_l}^2|{\bar
H_l}|^2+m_H^2|H|^2+\L{\ld_l A_{\ld_l}} H\bar H_l H_l+h.c.\R
\label{soft}.
\end{align}
Among the soft terms, $m^2_{\bar H_l}$ and $m^2_{ H_l}$ are positive
while $m_H^2$ is negative. $A_{\ld_l}$ is purely RGE induced,
roughly given by (see Eq.~(\ref{Aldl}))
\begin{align}\label{}
A_{\ld_l}\simeq &-{8 \over 16\pi^2}  {(Q_{H_l}g_X)^2 m_{\wt
X}}\log{M\over \mu_d}\sim -10^{-2}{\rm \,GeV}\label{Al},
\end{align}
which is much smaller than the typical scale in the dark sector
and thus is not a relevant soft parameter although it controls the
mixing between $H_l$ and $\bar H_l^*$. $H$ is the Higgs field
which breaks \U gauge symmetry. Its potential is simply a complex
$\phi^4$ theory, where the negative $m_{H}^2$ and quartic term
from $D-$term stabilizes the potential at the minimum
\begin{align}\label{vl}
\lan H\ran=v_{H}={|m_{H}|\over Q_Hg_X}\sim {\cal O}(10) \,{\rm GeV}.
\end{align}
The dark spectrum can be at the required several-GeV scale simply by
setting $\ld_l\sim 0.5$. The dark gauge
boson mass is  given by $m_X=\sqrt{2}Q_Hg_Xv_{H}=\sqrt{2}
|m_{H}|\simeq 0.2$ GeV, depending only on the negative Higgs
parameter.  To calculate the dark spectrum and the interactions in
the dark sector, we take a unitary gauge to eliminate the Goldstone
boson from the spectrum and write
 \begin{align}
 H=v_H+{{H}_R\over
\sqrt{2}}.
\end{align}
The CP-even state $H_{R}$ does not mix with other states and gets
its mass from the quartic term ($D-$term). Since the $D-$term is
determined by gauge coupling, at tree-level $H_{R}$ is exactly as
light as the dark gauge boson. The LDM can annihilate into such
light bosons too effectively and thus the freeze-out relic density
is too low, which will be discussed in the next section.

Now we study the states from the superfields $(\bar H_l,H_l)$.
The mass-square matrix of the complex scalars in the basis
of $( H_l^*,\bar H_l)$ is given by
\begin{align}\label{h}
M_{l}^2=\L\begin{array}{cc}
  \ld_l^2v_H^2+m_{ H_l}^2 & \ld_l A_{\ld_l} v_H \\
  \ld_l A_{\ld_l} v_H&  \ld_l^2v_H^2+m_{\bar H_l}^2
\end{array}\R.
\end{align}
Here we do not consider CP-violation and thus all parameters are
taken as real. The soft trilinear term is only a RGE effect and is
a small perturbation to the diagonal elements. So the two mass
eigenstates
 are approximately same as the interaction
states ($H_l^*, \bar H_l$), with the eigenvalues given by
the two diagonal elements in $M_{l}^2$. Because of the positivity of
$m_{H_l}^2$ and $m_{\bar H_l}^2$, these bosons are heavier
than the Dirac fermion formed by the two Weyl fermions
$\chi=(\wt H_l,\wt{ \bar H}_l)$, whose mass is given by
$M_L\equiv \ld_l v_H$. Such a \U-charged Dirac fermion
$\chi$ serves as the LDM candidate.

Note that in our framework we do not choose a singlet type dark
sector as in \cite{Cheung:2009qd}, where the dark sector
superpotential is $NH' H$, with $N$ being a singlet and
$Q_{H'}=-Q_{H_h}$,  and no $SU(N)$ flavor symmetry is introduced.
Since in that case $N$ is a singlet with $m_N^2<0$ at $\mu_d$
obtained from renormalization only, the LDM is always a
singlet-like scalar and its couplings with quarks are suppressed
by an extra mixing factor $1/\delta_A^2$, with $1/\delta_A$
measuring the fraction of the charged component $H'$ in the LDM,
given by
\begin{align}
\delta_A&\approx
 { m_{\bar H}^2-m_{N}^2\over |\ld_l A_{\ld_l} v_H|} \sim {2 \sqrt{2}\over\ld_lg_X}\sim {\cal O}(10^{2}).
\end{align}
In this estimation we assumed an ideal case that the pure \U
mediation contribution  to $m_{H}^2$ and $m_{\bar H}^2$ equals to
the renormalization contribution from ${\cal S}_X$. So in that
scenario, besides the suppression from $\theta^2\sim 10^{-10}$,
the LDM-nucleon scattering will be further suppressed by a factor
$1/\delta_A^4\sim 10^{-8}$, rendering the cross section
unacceptably small.

Finally, we comment on  $\wt H$ (fermionic component of $H$) and
dark gaugino $\wt X$. They have a Dirac mass $m_{HX}=m_X\simeq 0.2$
GeV. And $\wt X$ also has a heavier Majorana mass term $m_{\wt
X}\sim 0.5$ GeV for the choice $Q_{f_1}g_X\simeq 0.01$. This will
lead to a seesaw-like spectrum, $i.e.$, the lighter one is very
light, even as light as tens of MeV. Provided that the SUSY breaking
scale is high enough, saying $\sqrt{F}\sim 10^9$ GeV, this particle
will be the LSP (otherwise, we have to make sure that after
decoupling, it decays away before the BBN).

\subsection{Light Dark Matter Relic Density}

In this paper we mainly discuss the LDM phenomenology and try to
explain the CoGeNT and DAMA/LIBRA results together with other
null results from XENON100, XENON10 and CDMS(Si).
We use the latest  data  analysis in \cite{Hooper:2010uy}, which showed
that the combination of DAMA/LIBRA and CoGeNT data can be well accommodated
by a LDM with a mass of $\sim 7$ GeV and an elastic scattering cross section
with the nucleon of $\sim 2\times 10^{-40}$ cm$^2$. Moreover, it showed that
such a LDM is not excluded by other null results.

In our framework we have such a  LDM  from the dark sector, the
Dirac fermion $\chi$. However, a proper relic density for this LDM
is hard to obtain from the standard freeze-out thermal production.
In fact, there are two annihilation channels for this LDM: one is
directly to the dark gauge boson with a rate $\propto4\pi
(Q_{H_l}g_X)^4/m_L^2$ and the other is to $H_{R}$ with a rate
$\propto4\pi \ld_l^4/m_L^2$. Clearly, without the suppression of any
large mass scale (e.g., a weak scale heavy field in the propagator),
the only way to keep the LDM as a weakly interacting massive
particle (WIMP) with a typical weak reaction rate $\sigma_0\sim
10^{-8}$ GeV$^{-2}$ is to set $Q_{H_l}\,g_X, \ld_l\lesssim 0.03$.
But such a smaller $\ld_l$ implies $v_H$ must  take  several hundred
GeVs to keep the  mass of $\chi$ is about 7 GeV. Anyway, in
principle it is a viable solution, for example, by taking
\begin{equation}
Q_{H}g_X\simeq 0.001,\quad \ld_l\simeq 0.03,\quad
\theta \simeq 10^{-4},
\end{equation}
and keep $m_X$ as light as 0.2 GeV, $i.e.$, $m_{H}^2(\mu_d)\simeq
-0.04$ GeV$^2$. But from Eq.~(\ref{mhl}) we have to choose
$Q_{f_1}\,g_X\simeq Q_{{H_h}}\,g_X\simeq 0.2$.  Although this
solution  has a virtue that it allows the HDM to be the Sommerfeld
type instead of decaying HDM (since HDM couples to dark gauge
boson with a large strength), it is at the price of a surprisingly
large \U charge hierarchy between the different fields, $e.g.$,
$Q_{H_l}:Q_{H_h}=1:200$. So we propose that the LDM abundance is
produced by the late-decay of the ordinary next-to-the lightest
supersymmetric particle (ONLSP) in the NMSSM (the collider
phenomenology of such ONLSP decay to dark states is studied in
\cite{Baumgart:2009tn}). In this solution, $\lambda_l$ takes a
large value so that the LDM annihilates to $H_{R}$ very fast and
eventually leaves a small abundance after decoupling. However, in
the presence of a light \U sector, the ONLSP will dominantly decay
to the \U-charged dark states. If this decay happens after the
decoupling of the LDM (typically $\sim 10^{-5} s$), then the
number density of the ONLSP is transferred to the LDM. Of course,
a rather large number density is needed because the relic energy
density of the LDM is proportional to its mass. But if the ONLSP
has a very weak annihilation, its relic number density can be
quite large. In the following we discuss this issue
quantitatively.

First we consider the lightest neutralino $N_1$ as the ONLSP. If
$N_1$ dominantly decays to dark Higgsino and dark Higgs, its
lifetime is estimated to be
\begin{eqnarray}
\tau_{N_1\rw h+\wt h}&\sim& \L Q_{H_l}^2\alpha_X f_{\wt
B}^2\theta^2M_{N_1}\R^{-1} \nonumber \\ 
& \simeq & 7\times
10^{-14}\times\L{10^{-5}\over \theta }\R^2\L {10^{-3}\over
Q_{H_l}^2\alpha_X}\R \L{1\over f_{\wt B}}\R^2\L{100 {\rm \,GeV}\over
M_{N_1}}\R s,
\end{eqnarray}
where $f_{\wt B}$ is the fraction of bino in $N_1$.
In order for this decay to be late enough, the bino component should be
highly suppressed $\sim 10^{-4}$.

Then we assume the right-handed snuetrino (sRHN) as the ONLSP.
Such a sRHN is present in the NMSSM extended with a right-handed
neutrino, which was used to explain the light neutrino masses by
seesaw mechanism \cite{Grossman:1997is,ArkaniHamed:2000bq}. We
consider a simple model with only one flavor RHN (denoted as $N$)
and lepton doublet introduced. Its relevant superpotential is
given by
\begin{align}\label{}
W_N&=Y^N LH_uN+{M_N\over 2}N^2+{\ld_{SN}\over 2} SN^2+\mu H_uH_d,
\end{align}
where $\mu\equiv \ld v_s$. Depending on the $Z_3$-charge
assignment, $M_N$ or $\ld_{SN}$ can be turned off. In the
following we focus on the case with $\ld_{SN}=0$. Further, in GMSB
the soft terms involving the SM singlet $N$ can be dropped safely
because they are all generated by RGE effects from the coupling to
$L$ and $H_u$ (such renormalization effect is suppressed by
$Y^N\sim 10^{-5}$ in the low-scale seesaw). The LR-mixing is
naturally suppressed. In the CP-eigenstate basis of sleptons $(\wt
\nu_+^*,\wt N_+^*,\wt \nu_-^*,\wt
 N_-^*)$, the mass matrix is
\begin{align}\label{snu}
M_{\wt \ell}^2\approx \L\begin{array}{ccccc}
                           m_{\wt \ell}^2+D^2  & \quad\quad F^2+m_DM_N & 0  & 0 \\
                            &  m_{\wt N}^2+M_N^2+B_NM_N & 0 & 0  \\
                            &     & m_{\wt \ell}^2+D^2 & F^2-m_DM_N \\
                             &   &    & m_{\wt N}^2+M_N^2-B_NM_N
                         \end{array}\R,
\end{align}
where $D^2=0.5 m_Z^2\cos(2\beta)$ and  the mixing parameters
$F^2\approx -\mu m_D\cot \beta$   \cite{Arina:2007tm}. The lightest
state $\wt \nu_1$ with mass-square $m_{\wt \nu_1}$ is dominated
by $\wt N_-^*$, provided that
\begin{align}\label{}
m_{\wt \nu_1}^2\approx m_{\wt N}^2+M_N^2-B_NM_N\approx  M_N^2<
m_{\wt \ell}^2+D^2.
\end{align}
We have used the fact that the splitting is small, so the mass eigenvalues
are nearly the four diagonal elements. Concretely, the component of
$\wt \nu_-^*$ is given by $\wt \nu_1\supset {\cal C}_{1}^-\wt
\nu_-^*$ with
\begin{align}\label{}
{\cal C}_{1}^-\approx  {F^2-m_DM_N\over \delta m_{12}^2
}\approx{F^2-m_DM_N\over m_{\wt \ell}^2-M_N^2},
\end{align}
where $\delta m_{12}^2 $ is the mass-square splitting between the
two mass eigenstates of 3-4 block in the matrix Eq.~(\ref{snu}).
Depending on the mass splitting and $M_N$, the fraction covers  over  a
wide region:
\begin{align}\label{}
|{\cal C}_{1}^-|\simeq  {\sqrt{m_\nu M_N^3}\over \delta
m_{12}^2}\sim 10^{-8}-10^{-2} ,
\end{align}
where we used the seesaw formula for the light neutrino mass scale
$m_\nu=m_D^2/M_N\sim 0.1$ eV.

\begin{figure}[htb]
\begin{center}
\includegraphics[width=5.0in]{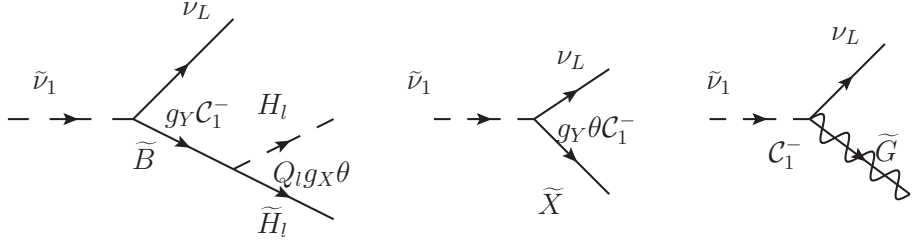}
\end{center}
\vspace*{-.7cm} \caption{\label{decay} The $\wt \nu_1$ decays to
dark states and gravitino.}
\end{figure}

As the  ONLSP, the $\wt \nu_1$ has two decay channels to dark sector
through its left-handed slepton component. One is the interesting
three-body decay via $\wt \nu_1\ra \nu_L+H_l+\wt H_l$ mediated by
bino, as shown in Fig.~\ref{decay}, and the decay lifetime is
\cite{Baumgart:2009tn}
\begin{align}\label{}
\tau_{\wt \nu_1}\sim& \L Q_{H_l}^2\alpha_X f_{\wt B}^2({\cal
C}_{1}^-)^2\theta^2{m_{\wt \nu_1}\over 16\pi^2P(m_{\wt
\nu_1}/M_{1})}\R^{-1}\cr \simeq& 2.6\times 10^{-3}s\times
\L{10^{-4}\over f_{\wt B}{\cal C}_{1}^-}\R^2 \L{10^{-5}\over \theta
}\R^2 {10^{-3}\over Q_{H_l}^2\alpha_X}   {300 {\rm \,GeV}\over
m_{\wt \nu_1}}{1\over P(m_{\wt \nu_1}/M_{1})}.
\end{align}
The other channel is $\wt \nu_1\ra \nu_L+ \wt X$ (also see
Fig.~\ref{decay}), but is suppressed by an additional helicity
factor $(m_{\wt X}/ M_1)^2$ and typically several times smaller than
the three-body decay \cite{Baumgart:2009tn}. Moreover, it can also
the decay into Goldstino $\wt \nu_1\ra \nu_L+ \wt G$, as shown in
Fig.~\ref{decay}, which is suppressed by the SUSY-breaking scale
$\sqrt{F}\gtrsim 10^3$ TeV. So, the $\wt \nu_1$ ONLSP mainly decays
to the \U charged dark states before BBN era ($\gtrsim 1s$) and thus
can provide a proper LDM density.

To end up this section, we point out one merit of the  LDM  from
late decay. In Ref.~\cite{Lavalle:2010yw} it was shown that if the LDM
reaches its relic density via annihilating to SM fermions, then the
required LDM-nucleon scattering cross section generally implies
antiproton excess, leading to some tension. But obviously our LDM
scenario evades this constraint.

\subsection{Explanation of CoGeNT and DAMA/LIBRA Results}
The study in \cite{Hooper:2010uy} suggests a $\sim 7$ GeV light dark matter
with an elastic scattering cross section with the nucleon of $\sim 2\times 10^{-40}$
cm$^2$. In the following we study the LDM interaction with the nucleon.

Using the  method  described  in \cite{Jungman:1995df}, we
derive the effective interaction between the LDM and the nucleus.
The microscopic interaction is presented in Appendix.~\ref{kin}.
Due to the kinetic mixing, the LDM interacts with
quarks, mediated by the dark gauge boson. It is the basis of the
effective theory describing the LDM-nucleon interaction. This effective
theory  is obtained by calculating  the quark and gluon operators in
a nucleon state, such as $\langle n|\bar f\gamma_\mu f|n\rangle$.
Then we obtain an interaction:
\begin{align}\label{LDM-nucleus}
{\cal L}_{vec}^{n,p}&= J^\mu_X\left(X_\mu+\theta s_W Z_\mu
\right)+f_n X_\mu \overline{ n} \gamma^\mu n+f_p X_\mu \overline{ p}
\gamma^\mu p, \cr
 J^\mu_X &= {1\over 2}(Q_{H_l}+Q_{\bar H_l}) g_X\bar \chi
 \gamma^\mu\chi+{1\over 2}(Q_{H_l}-Q_{\bar H_l}) g_X\bar \chi
 \gamma^\mu\gamma_5\chi\cr
 &+iQ_lg_X\L \partial^\mu H_l H_l^\dagger-\partial^\mu H_l^\dagger
 H_l
 \R+\cdots,
\end{align}
where the dots denote the irrelevant contributions like the
gauginos. From Eq.~(\ref{charge}) we get $Q_{H_l}+Q_{\bar
H_l}=Q_{H_h}$. And the term proportional to $|Q_{H_l}-Q_{\bar
H_l}|=\sqrt{\frac{4-N}{3N}}|Q_{H_h}|<|Q_{H_l}+Q_{\bar H_l}|$ induces
spin-dependent scattering. But it is suppressed by the smaller
charge $|Q_{H_l}-Q_{\bar H_l}|$. Moreover, for the nucleus with
large atomic number $A>20$, it is usually dominated by
spin-independent scattering \cite{Jungman:1995df}. Thus in the
following discussion we only keep the contribution from
spin-independent scattering. Note that the LDM-quark interaction is
mediated by the dark gauge boson. Consequently, due to the
conservation of the vector current, the sea quarks and the gluons
will not contribute to the current operator $\bar f\gamma_\mu f$. As
a result, the derivation of the effective theory is not only
simplified, but also free of uncertainty from considerations like
spin or strangeness content of the nucleon. This implies that the
effective $U(1)_X$ charge of the nucleon, $f_{p,n}$, only receives
contribution from its constituent quark \cite{Jungman:1995df}. So we
have
\begin{align}\label{}
f_n=b_u+2b_d,\quad f_p=2b_u+b_d,
\end{align}
where  $b_{u,d}=\theta g_Y\cos^2\theta_W Q_{u,d}$ with $Q_{u,d}$ being
the QED-charge of $u,d$ quarks. Thus at the leading order only the
QED-charged proton carries a tiny \U charge. The vector interaction
only mediates the spin-independent interaction between dark matter
and the nucleon and thus the LDM-nucleon scattering cross section can be
added coherently to give the total LDM-nucleus cross section.
That means that given the LDM-nucleon cross section $\sigma_p$,
the LDM-nucleus cross section is proportional to $ \L
Zf_p+(A-Z)f_n\R^2\sigma_p$ with  $Z$ and  $A$  being respectively
the proton and atomic numbers of the nucleus.

In practice, the four-fermion effective interaction is enough for
the calculation of the LDM-nucleus cross section since, for each
scattering by exchanging a dark gauge boson, the typical
transferred momentum is $(p_1-p_2)^2= |q|^2=2 \mu_N^2
v^2(1-\cos\theta)\sim {\cal O}(10^{-4})$GeV$^2\ll m_X^2\simeq
4\times 10^{-2}$ GeV$^2$. The reduced mass is
$\mu_N=m_Nm_L/(m_L+m_N)\simeq m_L$ for large nucleus like Ge with
mass  $m_N\approx73$ GeV. Thus the $X$ boson can be integrated
out.

From Eq.~(\ref{LDM-nucleus}) we calculate the scalar
spin-independent scattering cross section between the LDM and the
proton. It is given by
\begin{align}
\sigma_p\approx \f{N}{4}{\mu_p^2\over \pi }{(Q_Hg_X)^2\over
m_X^4}f_p^2,
 \end{align}
with $N$ being the internal index from $SU(N)$.
This result is valid only in the  non-relativistic limit (zero
momentum transfer). Note that in most previous studies the results
are usually displayed on the plane of DM mass versus the DM-nucleon
scattering cross section by setting $f_p=f_n$.
But in our model, $f_n\approx 0$, and thus $\sigma_p$ should be
re-scaled as $\sigma_p Z^2/A^2 \varpi$ when compared with data,
where $\varpi$ is the fraction of LDM in the total DM. Then
we have
\small
\begin{align}
\sigma_p\rightarrow &\left({Z^2\over A^2 \varpi}\right)\f{N}{4}
{m_p^2\over m_X^4 }{g_1^2Q_{H_h}^2g_X^2\over \pi }\theta^2
\cos\theta_W^4\cr \sim&
 \left({Z^2\over
A^2 \varpi}\right)\times 1.2\times 10^{-40} \times N\left({0.2
\,{\rm GeV}\over m_X }\right)^4 \left({Q_Hg_X \over
0.02}\right)^2\left({\theta \over 2\times 10^{-{5}} }\right)^2 {\rm
cm}^2 .
\end{align}
\normalsize
The cross section is independent of the LDM mass. Different
experiments have different values for the ratio $(Z/A)^2$.
For a LDM with a mass of $\sim 7$ GeV, this cross section is
just at the right order, according to the analysis in \cite{Hooper:2010uy}.

\section{Conclusion}
Both the cosmic ray anomalies observed by PAMELA and Fermi-LAT
experiments and the possible events from direct detections
like CoGeNT and CDMS II experiments
may indicate the existence of dark matters. But the former points to
a heavy dark matter at TeV sacle, while the later favors a light
dark matter with a mass of several GeV. Meanwhile, the Sommerfeld
enhancement may imply a dark \U gauge boson with a sub-GeV mass. In
light of these apparently different mass scales, we in this work
constructed a simple and coherent framework with GMSB, based on the
NMSSM extended with a light \U sector and a heavy dark matter
sector. By coupling the heavy dark matter directly to the \U-charged
messengers in the hidden sector, our framework has the following
intriguing features:
\begin{itemize}
\item[(1)] The kinetic mixing $\theta\sim10^{-5}$ is obtained
after integrating out the messengers with small doublet-triplet splitting.

\item[(2)] A large negative mass-square $m_S^2$ for the NMSSM singlet $S$
is generated at the messenger scale $M$.

\item[(3)] The dark \U is spontaneously broken at the GeV-scale, while the
dark gauge boson can have a sub-GeV mass by assuming a weakly \U-charged
Higgs field.

\item[(4)] A GeV-scale light dark matter with the required interaction
strength with quarks is provided. And its normal relic density can
be generated by the ONLSP late decay.
\end{itemize}

\begin{acknowledgments}
We would like to thank Ping He, Pengfei Yin and Yufeng Zhou  for useful
discussions on direct detection. This work was supported by the
National Natural Science Foundation of China under grant Nos.
10821504, 10725526 and 10635030, by the DOE grant
DE-FG03-95-Er-40917, and by the Mitchell-Heep Chair in High Energy
Physics.
\end{acknowledgments}

\appendix

\section {Kinetic Mixing and Dark-Visible Interaction}
\label{kin}
The messengers in the hidden sector are charged under
$U(1)_Y\times U(1)_X$. After the $SU(5)-$ gauge group is broken,
these fields generate the kinetic mixing between the two groups.
The effective theory (for the bosonic gauge part only,
with fermionic part obtained similarly) below the messenger
threshold is given by
\begin{align}
{\cal L}_{gauge}=-{1\over 4}F_Y^{\mu\nu}F_{Y\mu\nu}-{1\over
4}F_X^{\mu\nu}F_{X\mu\nu}+{\theta\over 2}F_Y^{\mu\nu}F_{X\mu\nu}.
\end{align}
Setting the initial value $\theta=0$ at $M_{GUT}$, then we get the
value at the  messenger scale $M$ through one-loop RGE
\cite{delAguila:1988jz}
\begin{align}
16\pi^2 {d\theta\over dt}\approx&2\sum_I g_Xg_1Q_Y^IQ_X^I,
\end{align}
where $I$ runs over all the superfields with $U(1)_Y\times U(1)_X$
charge. After electroweak symmetry breaking, the three gauge bosons
$Z'_\mu$, $B'_\mu$ and $X'_\mu$ mix with each other through kinematic
terms and mass terms.
Using the convention in \cite{Baumgart:2009tn}, after eliminating
the mixing and working in the mass eigenstate  basis
$(Z_\mu,A_\mu,X_\mu)$, the interactions between the gauge boson and
the current at the leading order of $\theta$ are described by
\small
\begin{align}
 {\cal L}_{coupling}&\supset \theta X_\mu \L\cos\theta_WJ^\mu_{em}+{\cal
O}(m_X^2/m_Z^2)J_Z^\mu\R 
+\theta Z_\mu\L-\sin\theta_W J_X^\mu+{\cal
O}(m_X^2/m_Z^2)J_W^\mu\R\cr &+ \theta \L \wt B \wt J_X
+{\cal
O}(m_{\wt X}/M_1) \wt X \wt J_B\R,
\end{align}
\normalsize
where the bosonic gauge  current $J_{em,W,Z}^\mu$ and $J_X^\mu$
are defined as usual, while its fermonic counterpart, $i.e.$, the
supercurrents, are defined as  $\wt J_X=g_X\sum_iQ_{X,i}\wt
d_i^\dagger d_i,\,\wt J_B=g_Y\sum_iQ_{Y,i}\wt v_i^\dagger v_i$
with $d_i/v_i$ denoting any dark/visible fermions. The $X_\mu
J^\mu_{em}$ accounts for the cosmic ray anomaly after HDM
decays/annihilates to the dark sector. And $\wt B \wt J_X$
provides the $U(1)_Y$ gaugino interaction with the dark states.

\section {Soft Terms from HDM-Messenger Direct Couplings}
If some fields feel the SUSY-breaking via direct couplings to
the messengers, they  will give new contribution to soft terms
controlled by Yukawa couplings. The soft terms can be extracted
by using wave-function renormalization method \cite{Giudice:1997ni}.
According to this method,
we need to know the discontinuity of anomalous dimensions
and the beta function across the messenger threshold
\cite{Delgado:2007rz}. Above the messenger scale we have
\begin{align}
\gamma_{H_h}=&{1\over 16\pi^2}\L -2\ld_h^2+4g_X^2Q_{H_h}^2\R,\cr
\gamma_{\bar H_h}=&{1\over 16\pi^2}\L
-2\ld_h^2-6\ld_T^2-4\ld_D^2+4g_X^2Q_{H_h}^2\R,\cr \gamma_S=&{1\over
16\pi^2}\L
-4\kappa^2-2\ld_h^2-2\ld^2\R,\\
\beta_{\ld_h}=&{2\ld_h^2\over 16\pi^2}\L
2\kappa^2+3\ld_h^2+\ld^2+3\ld_T^2+2\ld_D^2-4g_X^2Q_{H_h}^2\R,\cr
\beta_{\ld_T}=&{2\ld_T^2\over 16\pi^2}\L
5\ld_T^2+\ld_h^2+2\ld_D^2-2g_X^2(Q_{f_1}^2+Q_{H_h}^2)-\f{4}{9}g_1^2-\f{16}{3}g_3^2\R,\cr
\beta_{\ld_D}=&{2\ld_D^2\over 16\pi^2}\L
3\ld_T^2+\ld_h^2+4\ld_D^2-2g_X^2(Q_{f_1}^2+Q_{H_h}^2)-g_1^2-
{3}g_2^2\R.
\end{align}
 In the calculation, $X$ is taken as a non-propagating background.
Across the messenger threshold, the various discontinuity is given by
\begin{align}
\Delta\gamma_{H_h}=&\,0,\quad  \Delta\gamma_{\bar H_h}=-{2\over
16\pi^2}(3\ld_T^2+2\ld_D^2),\quad \Delta\gamma_{S}=\,0, \cr \Delta
\beta_{\ld_h}=&{2\ld_h^2\over 16\pi^2}(3\ld_T^2+2\ld_D^2),\quad
\Delta\beta_\ld=\Delta\beta_\kappa=0,\cr
\Delta\beta_{g_X^2}=&10\f{2}{16\pi^2}Q^2_{f_1}g_X^4,\quad
\Delta\beta_{g_i}=c_in\f{g_i^4}{16\pi^2},
\end{align}
where $c=(5/3,1,1)$ and $n$ is the pair of $(5,\bar 5)$ messengers.
Using this result, we calculate the following soft terms at the
messenger boundary
\small
\begin{align}\label{mS}
m_S^2=&-\f{\ld_h^2}{(16\pi^2)^2}\L3\ld_T^2+2\ld_D^2\R\f{F^2}{M^2},\\\label{Aldh}
A_{\ld_h}=&-\f{1}{16\pi^2}\L3\ld_T^2+2\ld_D^2\R \f{F
}{M},\\\label{Hh}
 m_{H_h}^2=&m_S^2+m_G^2,\\\label{bHh}
  m_{\bar H_h}^2=&\f
{1}{(16\pi^2)^2}\left[8\ld_D^4+15\ld_T^4+12\ld_T^2\ld_D^2-16g_3^2\ld_T^2
 \right. \nonumber \\
 &\left.
-6g_2^2\ld_D^2-2g_1^2\L\f{2}{3}\ld_T^2+\ld_D^2\R-2g_X^2(Q_{f_1}^2+Q_{H_h}^2)(3\ld_T^2+2\ld_D^2)
\right]\f{F^2}{M^2}+m_{G}^2,
\end{align}
\normalsize
where $m_G^2= (Q_{H_h}/Q_{f_1})^2 m_{\wt X}^2/(2n_X)$ (with dark
gaugino mass $m_{\wt X}=2n_X{Q_{f_1}^2g_X^2\over 16\pi^2}{F\over M}$
and $n_X=5$ in our paper) is the pure \U GMSB contribution. And
similar formula applies to the gauge mediation contribution to the
soft mass term of other fields by  replacing  $Q_{H_h}$ with
corresponding charge.

\section {One-Loop RGEs of Some Soft Terms}
Here we present some important one-loop RGEs for all the soft terms in the dark sector,
$m_S^2$, $A_\ld$ and $A_\kappa$ in the NMSSM,  and some Yukawa couplings and dark
gauge couplings. In general, they take the form:
\begin{align}
{dY\over dt}={1\over 16\pi^2}\beta_Y,\quad {dA\over dt}={1\over
16\pi^2}\beta_A, \quad {dm^2\over dt}={1\over 16\pi^2}\beta_{m^2},
\end{align}
where $t\equiv \log (Q/Q_0)$ with $Q_0$ being the boundary energy scale and $Q$ the
running scale.
 Following a general calculation in \cite{Martin:1993zk}, the RGEs for the
soft terms in the dark sector are given by
\small
\begin{eqnarray}\label{Aldl}
\beta_{A_{\ld_l}}&=&9\lambda_l^2
A_{\ld_l}+g_X^2Q_{H_l}^2\left(8m_{\wt
X}\lambda_l-4A_{\ld_l}\right),\\\label{mH2}
 \beta_{m_H^2}&=&2N
\lambda_l^2\left(m_{H_l}^2+m_{\bar
H_l}^2+m_H^2+A_{\ld_l}^2\right)-8g_X^2Q_{H}^2m_{\wt
X}^2+2Q_Hg_X^2{\cal S}_X,\\\label{mHl}
\beta_{m_{H_l}^2}&=&2\lambda_l^2\left(m_{H_1}^2+m_{\bar
H_1}^2+m_H^2+A_{\ld_l}^2\right)-8g_X^2Q_{H_l}^2m_{\wt
X}^2+2Q_{H_l}g_X^2{\cal S}_X,\\\label{mbHl2}
 \beta_{m_{\bar
H_l}^2}&=&2\lambda_l^2\left(m_{H_1}^2+m_{\bar
H_1}^2+m_H^2+A_{\ld_l}^2\right)-8g_X^2Q_{\bar H_l}^2m_{\wt X}^2+2
Q_{\bar H_l}g_X^2{\cal S}_X,
\end{eqnarray}
where ${\cal S}_X$ is defined in the text.  For the
modified NMSSM soft parameters, the new RGEs are given by
\small
\begin{eqnarray}\label{Aldl}
\beta_{m_S^2}&=& 4\ld^2\L
m_{H_u}^2+m_{H_d}^2+m_S^2+A_\ld^2\R+4\kappa^2\L
3m_S^2+A_\kappa^2\R\cr &&+2\ld_h^2\L m_S^2+m_{H_h}^2+m_{\bar
H_h}^2+A_{\ld_h}^2\R,
\\\label{AldR}
\beta_{A_{\ld}}&=&8\ld^2A_\ld+6h_t^2A_t+6h_b^2A_b+2h_\tau^2A_\tau+4
\kappa^2A_\kappa+2g_1^2M_1+6g_2^2M_2 
+2\ld_h^2A_{\ld_h} \\ \label{Akappa} 
\beta_{A_{\kappa}}&=&
12\kappa^2A_\kappa+12\ld^2A_{\ld}+2\ld_h^2A_{\ld_h},
\end{eqnarray}
\normalsize
Finally, for the new Yukawa couplings and the dark gauge coupling, their
RGEs are given by
\begin{eqnarray}\label{ldh}
\beta_{\ld_h}&=&2\ld_h^2\L2\kappa^2+3\ld_h^2+\ld^2-4(Q_{H_h}g_X)^2\R\quad({\rm
below \,\,}M),\\
\beta_{\ld_l}&=&2\ld_l^2\L3\ld_l^2-4(Q_{H_h}g_X)^2\R \quad({\rm
below
\,\,}M),\\
\label{gx}
 \beta_{g_X^2}&=&2g_X^4\L10Q_{f_1}^2+2Q_{H_h}^2+Q_{H_h}^2+N(Q_{\bar H_l}^2+Q_{H_l}^2)\R \quad({\rm
above \,\,}M),
\\
\label{xiT}
 \beta_{\xi_{1,T}}&=&\xi_{1,T}\L2\xi_{1,T}^2+\ld_T^2-\f{16}{3}g_3^2-\f{9}{15}g_1^2\R \quad({\rm
above \,\,}M),
\\
\label{xiD}
 \beta_{\xi_{1,D}}&=&\xi_{1,D}\L2\xi_{1,D}^2+\ld_D^2-{3}g_2^2-\f{4}{15}g_1^2\R
\quad({\rm
above \,\,}M).
\end{eqnarray}
At the scale $M_{GUT}$, the unified value $\xi_{1,D}=\xi_{1,T}$ is assumed.

\end{document}